\documentclass[12pt]{article}
\usepackage{amsfonts,amssymb,epsfig,amsmath,times}
\usepackage{float}
\restylefloat{table}
\renewcommand{\text}[1]{#1}

\newcommand{\be}{\begin{equation}}
\newcommand{\ee}{\end{equation}}
\newcommand{\ben}{\begin{displaymath}}
\newcommand{\een}{\end{displaymath}}
\newcommand{\bea}{\begin{eqnarray}}
\newcommand{\eea}{\end{eqnarray}}
\newcommand{\bean}{\begin{eqnarray*}}
\newcommand{\eean}{\end{eqnarray*}}
\newcommand{\nn}{\nonumber \\}
\newcommand{\ba}{\begin{array}}
\newcommand{\ea}{\end{array}}
\newcommand{\bi}{\begin{itemize}}
\newcommand{\ei}{\end{itemize}}

\def\G{\Gamma}

\def\e{\epsilon}

\def\otaula{\begin{tabular}}
\def\ctaula{\end{tabular}}

\def\bnum{\begin{enumerate}}
\def\enum{\end{enumerate}}

\def\CR{\mathcal{R}}
\def\CM{\mathcal{M}}

\def\8M{$\CM_8$}

\def\be{\begin{equation}}
\def\ee{\end{equation}}
\def\G{\Gamma}

\def\ei{e^{\underline{i}}}

\def\e1{e^{\underline{1}}}
\def\1u{\underline{1}}
\def\2u{\underline{2}}

\def\0u{\underline{0}}
\def\e{\epsilon}
\def\target{$\CR^{1,1}\times \mathcal{M}_8$ }
\def\target2{$\CR^{1,1}\times \mathcal{M}_8$,}
\def\9G{\G_{\underline{9}}}

\def\1f{f_1^{1/2}}
\def\2f{f_2^{1/2}}
\def\4f{f_4^{1/2}}

\begin{document}
\begin{titlepage}

\vfill

\begin{flushright}
KIAS-P09015
\end{flushright}

\vfill

\begin{center}
   \baselineskip=16pt
   {\Large\bf Positronium-like states from supergravity}
   \vskip 2cm
      Eoin \'{O} Colg\'{a}in and Hossein Yavartanoo
         \vskip .6cm
      \begin{small}
      $^1$\textit{Korea Institute for Advanced Study, \\
        Seoul 130-722, Republic of Korea}
        \end{small}\\*[.6cm]
\end{center}

\vfill
\begin{center}
\textbf{Abstract}\end{center}
\begin{quote}
We study semi-classical rotating open strings ending on a probe Dp-brane close to a stack of extremal, non-extremal Dp-branes and Dp-branes with $B$-fields. These states may be interpreted as positronium-like bound states of spinning W-bosons in an $SU(N+1) \rightarrow SU(N) \times U(1)$ theory. In each case, we describe the behaviour of the strings and examine the relation between the energy and the angular momentum of the bound state.
\end{quote}

\vfill

\end{titlepage}

\section{Introduction}
The quark anti-quark potential, determined from the expectation value of Wilson loops, is an important physical observable that earmarks the onset of confining phases in gauge theories. Following the advent of the AdS/CFT correspondence, the quark anti-quark potential was determined in the context of $AdS_{5} \times S_{5}$ \cite{Rey:1998ik} by studying fundamental strings attached to D3-branes. 

In the initial work \cite{Rey:1998ik} the basic idea was to introduce a D3-brane at some fixed radius in $AdS_{5}$ spacetime, and to interpret it as corresponding to the dual $\mathcal{N} =4$ super-Yang-Mills with gauge group $SU(N+1)$ broken to $SU(N) \times U(1)$ by the finite separation of one D3-brane from the $N$ D3-branes sourcing the $AdS_5 \times S_5$. In this setting a fundamental string ending on the D3, but stretching to the $AdS$ boundary, corresponds to a W-boson charged under both these groups. The mass of such a state corresponds to the location of the D3 and in the infinite mass limit, close to the $AdS_5$ boundary, the authors of \cite{Rey:1998ik} determined the quark anti-quark potential at large $N$ and large 't Hooft coupling $\lambda = g_{YM}^2 N$. 

In a recent work \cite{Haque:2009hz} the dynamics of a W-boson and anti-W-boson pair of $SU(N+1) \rightarrow SU(N) \times U(1)$ gauge theory at finite angular momentum and finite mass were analysed. Although W and anti-W are attractive, the pair may be stabilised by adding orbital angular momentum so that the system resembles a positronium-like state. This work considers rotating strings ending on D3-brane in an $AdS_5 \times S_5$ background. In the semi-classical picture these strings capture the dynamics of the dual theory in the large $N$ and large $\lambda$ limit \cite{Gubser:2002tv}. The dynamics of this positronium system has some overlap with the analysis of meson spectra with flavour appearing in \cite{Kruczenski:2003be,Arean:2005ar}. In \cite{Kruczenski:2003be}, the meson spectrum of $\mathcal{N}=2$ SYM with fundamental matter was determined by considering rotating open strings attached to a D7-brane probe in $AdS_{5} \times S^{5}$. In \cite{Arean:2005ar} a similar embedding of D7-branes allows one to study the corresponding non-commutative gauge theory. Although we rely primarily on the above references, for other works involving meson spectra in supergravity backgrounds, see \cite{Paredes:2004is,Peeters:2005fq}. 

In this work, we extend the work of \cite{Haque:2009hz} by considering the extension of the $p=3$ i.e. D3-brane in $AdS_5 \times S_5$ case to include field theories in different dimensions living on the Dp-branes. As Dp-branes are BPS objects breaking half the supersymmetry, the low-energy worldvolume theory of a stack of $N$ Dp-branes preserves 16 supercharges. The $R$-symmetry group is $SO(9-p)$ and the coupling constant is $g_{YM}^2 = (2 \pi)^{p-2} g_s \alpha'^{(p-3)/2}$, thus determining the phase of the gauge theory and the appropriate description. The dual field theories are all at zero temperature $T=0$, but by keeping the energy density above extremality finite, one may consider non-supersymmetric theories at finite temperature. Finally, by introducing a non-commutativity parameter in the form of a $B$-field to the Dp-brane background before taking the appropriate limit \cite{Maldacena:1999mh}, one may also study non-commutative theories preserving sixteen supersymmetries. \footnote{Similar studies of these non-commutative field theories have also appeared in \cite{Cai:2000hn}.} 

\section{Spinning open strings in Dp-brane backgrounds}
In this section we introduce the configuration of rotating open strings presented in \cite{Haque:2009hz} to describe a bound state of W/anti-W pair. This has a similar setup to the rotated folded strings appearing in \cite{Gubser:2002tv}. The rotating strings end on a Dp-brane located at small finite separation from a stack of $N$ Dp-branes. This separation introduces a scale to the theory and reduces the gauge group of the dual field theory from $SU(N+1)$ to $SU(N) \times U(1)$. The Dp-brane is also parallel to the stack, so the backgrounds preserve half the supersymmetry, unless a finite temperature is introduced.

As the Dp-brane is close to the stack, it sees the near-horizon of the brane. The backgrounds we consider in this note follow. We will use $\rho$ and $\theta$ to denote the plane of rotation of the string. 

\textbf{Extremal Dp-brane:} \\
When the stack is comprised of extremal Dp-branes, the metric for this configuration is:
\bea
\label{met1}
ds^2 &=& \alpha' \left[ \left(\frac{u}{R}\right)^{\frac{7-p}{2}}(\mathcal{M}^{1,p-2} + d \rho^2 + \rho^2 d \theta^2 ) + \left( \frac{R}{u} \right)^{\frac{7-p}{2}} (du^2 +  u^2 d \Omega^2_{8-p}) \right], \nn
R^{7-p} &=& c_{p} g^{2}_{YM} N, \quad c_{p} = 2^{7-2p} \pi^{\frac{9-3p}{2}} \G(\frac{7-p}{2}),
\eea
where $\mathcal{M}^{1,p-2}$ denotes Minkowski spacetime in $d = p-1$,  $R$ is a length scale (for $p=3$ it is the $AdS$ radius), $u = r/\alpha'$ the energy coordinate is related to the radial coordinate, and the coupling of the dual field theory may be written in terms of the string coupling $g_s$ as $g_{YM}^2 = (2 \pi)^{p-2} g_s \alpha'^{(p-3)/2}$. There is also a dilaton and a five-form field strength, which we ignore as they do not appear in the analysis. 

\textbf{Non-extremal Dp-brane:} \\
Here the metric is similar to the metric above, but one may introduce a function $f(u)$ describing the non-extremality
\bea
\label{met2}
ds^2&=& \alpha' \biggl[ \left(\frac{u}{R}\right)^{{7-p\over2}}\left(-f(u) dt^2  + d \rho^2 + \rho^2 d \theta^2 +\cdots dx_{p}^2
\right) \nn
&+&\left(\frac{R}{u}\right)^{{7-p\over 2}}\left(\frac{du^2}{f(u)}
+u^2d\Omega_{8-p}^2
\right) \biggr],
\eea
where the deviation from extremality is governed by 
\be
f(u) = 1-(\frac{u_T}{u})^{7-p}. 
\ee
Again $R$ is unchanged and there is a horizon at $u = u_T$. Regularity of the Euclidean section obtained from $t \rightarrow i t_{E}$ requires that $t_{E}$ be identified with period
\be
\beta = \frac{1}{T} = \frac{4 \pi R}{7-p} \left( \frac{R}{u_T} \right)^{\frac{5-p}{2}}. 
\ee
This gives the temperature of the dual $p+1$-dimensional field theory. 

\textbf{Dp-brane with $B$-field} \\
Finally we consider the introduction of a $B$-field, and the associated non-commutativity parameter $b$. The metric with a  $B$-field turned on is \cite{Maldacena:1999mh}:
\bea
\label{met3}
ds^2&=& \alpha' \biggl[ \left(\frac{u}{R}\right)^{{7-p\over2}}\left(-dt^2 +  h(d \rho^2 + \rho^2 d \theta^2)
+\cdots+dx_{p}^2\right) \nn &+& \left(\frac{R}{u}\right )^{{7-p\over2}}(du^2+u^2d\Omega_{8-p}^2), \biggr]\nn
B_{\rho \theta}&=&\frac{\alpha'}{b}\frac{a^{7-p}u^{7-p}}{1+a^{7-p}u^{7-p}} \rho,
\quad  h=\frac{1}{1+a^{7-p}u^{7-p}},\nn
R^{7-p} &=&c_p{ g}^2_{YM}N b,\quad
a^{7-p}=\frac{b^2}{R^{7-p}}, \quad
{ g}^2_{YM}=(2\pi)^{(p-2)}  g_s \alpha'^{(p-3-2)/2},  
\label{NONMET}
\eea
where the parameter $b$ determines the value of the $B$-fields and also serves as a non-commutativity parameter for the dual gauge theory. Setting $b=0$ will bring us back to the original Dp-brane near-horizon. 

\subsection{String worldsheet}
In describing the dynamics of the open string ending on the Dp-brane, we will make use of the Nambu-Goto action 
\bea
S_{NG} &=& - \frac{1}{2 \pi \alpha'} \int d \tau d \sigma \left\{ \sqrt{(\dot{X} \cdot X')^2 - \dot{X}^2 X'^2} -  \mathcal{P}[ B] \right\}, 
\eea
where $\mathcal{P}[B]$ denotes the pull-back of the $B$-field to the worldsheet of the string $\Sigma$, dots and primes denote derivatives with respect to $\tau$ and $\sigma$ respectively, and the dot products are taken with respect to background metric. The string worldsheet is parameterised by $(\tau, \sigma)$ and the embedding is determined by the target space coordinates $X^{M}(\sigma, \tau)$.

Throughout this note we will make use of the following ansatz:
\be
t = \tau, \quad \theta = \omega \tau, \quad \rho = \rho(\sigma), \quad u = u (\sigma),
\ee
where $\omega$ is the constant angular velocity of the string. Later, from the explicit actions, one will see that the Lagrangian density $\mathcal{L}$ does not depend explicitly on $t$ and $\theta$, therefore the system has two conserved quantities: the energy $E$ and the angular momentum $J$, 
\bea
\label{cons_quants}
E = \omega \partial_{\omega} \mathcal{L} - \mathcal{L}, \quad
J = \partial_{\omega} \mathcal{L}. 
\eea

Although the equations of motion may be determined from the action, they must be supplemented with the boundary conditions 
\be
\label{bdyconds}
\frac{\partial{L}}{\partial(X')^{M}} \delta X^{M} |_{\partial \Sigma} = 0, 
\ee
to ensure that the action is stationary. Here $\delta u |_{\partial \Sigma} = 0$ as the endpoints of the string are attached to the brane, but $ \delta \rho |_{\partial \Sigma}$ is arbitrary, so we must impose 
\be
\label{rhobdy}
\frac{\partial L}{\partial \rho'}|_{\partial \Sigma} = 0. 
\ee
We will see later that this implies Neumann boundary conditions - the string ending orthogonally to the brane - for Dp-branes, whereas when there is a $B$-field, the string ends at an angle which depends on $b$.

\section{Strings ending on Dp-branes}
The parameters appearing in this system are $\omega, b, u_{T}, R$ and $u_{0}$ the position of the probe Dp-brane on which the string ends. In general, this problem may only be tackled numerically, but there are limits of small and large $\omega$ that allow analytic descriptions. For large $\omega \rightarrow \infty$, the length of the rotating string $\rho$ is very small $\rho \sim 1/\omega$ and behaves like a string rotating in flat spacetime \cite{Gubser:2002tv}. Here the energy $E$ is related to the angular momentum $J$ by $E \sim \sqrt{J}$, accounting for the constant leading Regge trajectory in all our graphs.

Although it is also possible to explore the small $\omega$ region analytically \cite{Kruczenski:2003be}, when the parameters $b$ and $u_{T}$ are non-zero, this becomes difficult. As a result, in this paper we focus on the numerical results and the plotted graphs to infer the behaviour. For simplicity we consider $R=1$ in solving the equations of motion.  
\subsection{Extremal case}
For this background (\ref{met1}), the string worldsheet action becomes
\be
\label{act1}
S = -\frac{1}{2 \pi} \int d \tau d \sigma \left(\frac{u}{R}\right)^{\frac{7-p}{2}} \sqrt{(1-\omega^2 \rho^2)[ \rho'^2 + \left(\frac{R}{u}\right)^{7-p} u'^2 ]}.
\ee
One may calculate the energy and the angular momentum:
\bea
E &=&  \frac{1}{2 \pi} \int d \sigma \left(\frac{u}{R}\right)^{\frac{7-p}{2}} \sqrt{\frac{\rho'^2 + \left(\frac{R}{u}\right)^{7-p} u'^2}{1-\omega^2 \rho^2}}, \nn
J  &=&  \frac{\omega}{2 \pi} \int  d \sigma  \left(\frac{u}{R}\right)^{\frac{7-p}{2}} \rho^2 \sqrt{\frac{\rho'^2 + \left(\frac{R}{u}\right)^{7-p} u'^2}{1-\omega^2 \rho^2}}.
\eea
As explained earlier, the equations of motion from the above action also require that boundary conditions (\ref{bdyconds}) are imposed. The string length is parameterised by $0 \leq \sigma \leq \sigma_{max}$, where $\sigma_{max}$ is an as of yet undetermined value corresponding to one of the endpoints of the string. 

Without loss of generality, we take the position of the Dp-brane to be unity: $u(0) = u(\sigma_{max}) = 1$. The other boundary condition (\ref{rhobdy}) means 
\be
{u}^{\frac{7-p}{p-5}} \rho' \sqrt{\frac{1- \omega^2 \rho^2}{\rho'^2 + u^{p-7} (u')^2}} |_{\partial \Sigma} = 0. 
\ee
for this background. This means either $\rho'|_{\partial \Sigma}=0$ with the string ending orthogonally on the Dp-brane, or $\omega \rho |_{\partial \Sigma} = 1$, which means that the endpoints of the string move at the speed of light. As pointed out in \cite{Kruczenski:2003be}, only the former permits a description to a bound state of W-bosons because the latter choice leads to a string with only one endpoint on the Dp-brane, a fact that was transparent from the numerics: only certain boundary conditions allow the string to return to the Dp-brane at $u(\sigma_{max}) = 1$. So we will be solving the equations of motion subject to Dirichlet boundary conditions for $u(\sigma)$ and Neumann boundary conditions for $\rho(\sigma)$:
\begin{figure}
\label{P2config}
 \begin{center}
  \includegraphics[scale=1]{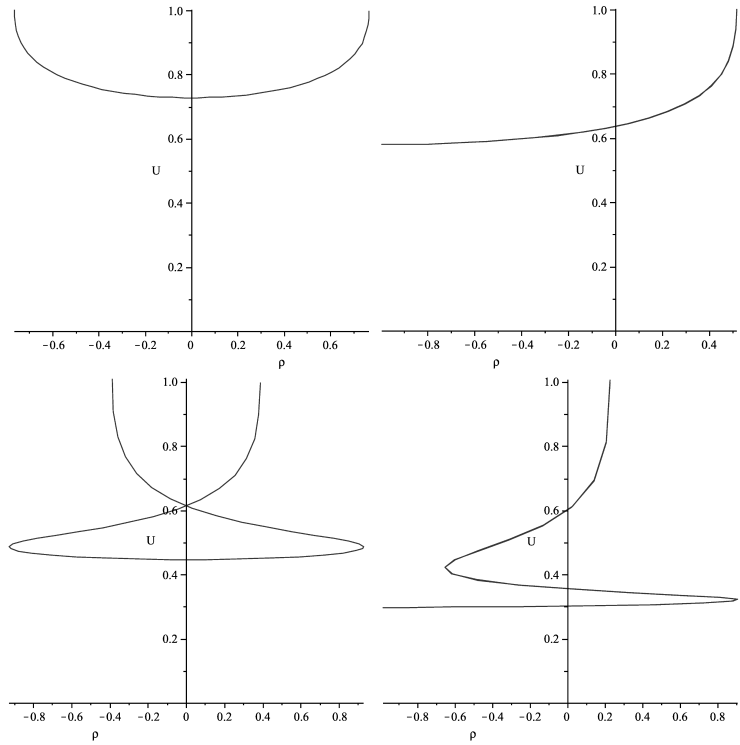}
 \end{center}
\caption{Various modes of the string for different nodes $n$ in case $p=2$. Apart from $n=0$ ``U'' shape, one sees folded strings and self-intersecting strings. Also worth noting is the depth increases as $n$ increases.}
\end{figure}
\be
\label{bdy1}
u(0 ) = u(\sigma_{max}) = 1, \quad \rho'(0) = \rho'(\sigma_{max}) = 0. 
\ee
In practice we fix $u(0) = 1$, while solving the equations subject to the constraints imposed by the reparameterisation invariance of the string worldsheet action. Here, for fixed $\omega$, $\rho (0)$ and $\sigma_{max}$ play the role of adjustable parameters that when tuned correctly allow the above constraints to be simultaneously satisfied. Starting from $\rho(0) = 1/\omega$, we identify the initial $n=0$ node and higher self-intersecting and folded ``excited '' string configurations, some of which are presented for $p=2$ in Fig. (\ref{P2config}). Similar configurations were noted for $p>2$, with the $p=3$ appearing in \cite{Haque:2009hz}.

Although considerably spaced at the beginning, the frequency of the permitted string configurations increases with decreasing $\rho(0)$. This is suggestive of the picture where the dual W-bosons are surrounded by shells of gluons associated to pieces of the string between each of the nodes. The closer the nodes are to the centre of the geometry, the larger the radius of the shell. 

\begin{figure}
\label{P2ext}
 \begin{center}
  \includegraphics[scale=.4]{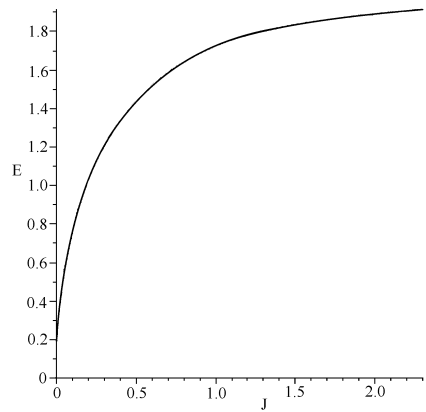}
 \end{center}
\caption{$E$ v. $J$ for extremal $p=2$.}
\end{figure}

As stated earlier, for $\omega >> 1$, the string is small and behaves as if it was in free space. However as $\omega \rightarrow 0$, the size of the string increases seeing more of the geometry, and one may ask what happens to the Regge behaviour as $\omega$ decreases. In Fig. (\ref{P2ext}) we plot $E(\omega)$ versus $J(\omega)$. When the string is large, its mass is just the mass of the W-bosons at its endpoints $E = 2 m_{W}$, thus accounting for the asymptotic behaviour.

\subsection{Dp-brane with $B$-field} 
In this section, we consider the generalisation of the previous background where $B$-fields are switched on. Similar analysis appeared for probe D7-branes in D3-brane with $B$-field backgrounds in \cite{Arean:2005ar}. The dynamics again is governed by the standard Nambu-Goto action, but this time the $B$-field of the background plays a role when it is pulled back to the worldsheet. The action is then
\bea
\label{act3}
S &=& -\frac{1}{2 \pi} \int d \tau d \sigma  \left(\frac{u}{R}\right)^{\frac{7-p}{2}} \sqrt{(1- h \omega^2 \rho^2)[ h \rho'^2 + \left(\frac{R}{u}\right)^{7-p} u'^2] } \nn &-& \frac{1}{2 \pi} \int d \tau d \sigma  \frac{1}{b} a^{7-p} u^{7-p} h \rho \omega \rho'. 
\eea
The conserved quantities in this case are 
\bea
E &=&  \frac{1}{2 \pi} \int  d \sigma  \left(\frac{u}{R}\right)^{\frac{7-p}{2}} \sqrt{\frac{h \rho'^2 + \left(\frac{R}{u}\right)^{7-p} u'^2}{1-h \omega^2 \rho^2}}, \nn
J &=& 
\frac{\omega}{2 \pi} \int  d \sigma \left(\frac{u}{R}\right)^{\frac{7-p}{2}} h \rho^2 \sqrt{\frac{\rho'^2 + \left(\frac{R}{u}\right)^{7-p} u'^2}{1-h \omega^2 \rho^2}}
- \frac{1}{2 \pi} \int  d \sigma \frac{1}{b} a^{7-p} u^{7-p} h \rho \rho' ,
\eea 
where the first expression is the physical angular momentum of the rotating string. 

We now turn attention to the boundary conditions. As before we have Dirichlet boundary conditions i.e. $u(0) = u (\sigma_{max})  = 1$,  so we only have to impose $\partial L/\partial \rho' |_{\partial \Sigma} = 0$. Taking into account the explicit form of the action, this entails
\be
\rho' \sqrt{\frac{1-h \omega^2 \rho^2}{ h \rho'^2 + u^{p-7}u'^2}} |_{\partial \Sigma} = -\omega \rho b u^{\frac{7-p}{2}} |_{\partial \Sigma},
\ee

\begin{figure}
\label{P2b}
 \begin{center}
  \includegraphics[scale=.5]{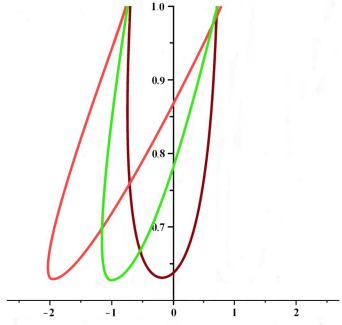}
 \end{center}
\caption{Tilting of $n=0$ mode for $b=1,5,10$.}
\end{figure}

where we have used $a^{7-p} = b^2/R^{7-p} = b^2$ for $R=1$. 
It is worth noting here that when the $B$-field (non-commutativity parameter) is zero $b=0$, we again recover Neumann boundary conditions. Instead, we see that the string ends on the Dp-brane at an angle which depends on $b$ and also on the value of $\rho$. 

This equation determines the angle at which the string touches the Dp-brane. Notice that when the $B$-field is set to zero, the latter term vanishes and we return to the boundary conditions (\ref{bdy1}) in the previous section. The actual value of $d u / d \rho$ at the ends of the string may be determined by solving the above quadratic. One obtains
\be
\frac{d u}{d \rho} |_{\partial \Sigma} = \pm \frac{\sqrt{1-\omega^2 \rho^2}}{b \omega \rho} |_{\partial \Sigma}. 
\ee

\begin{figure}
\label{EJNC}
 \begin{center}
  \includegraphics[scale=.35]{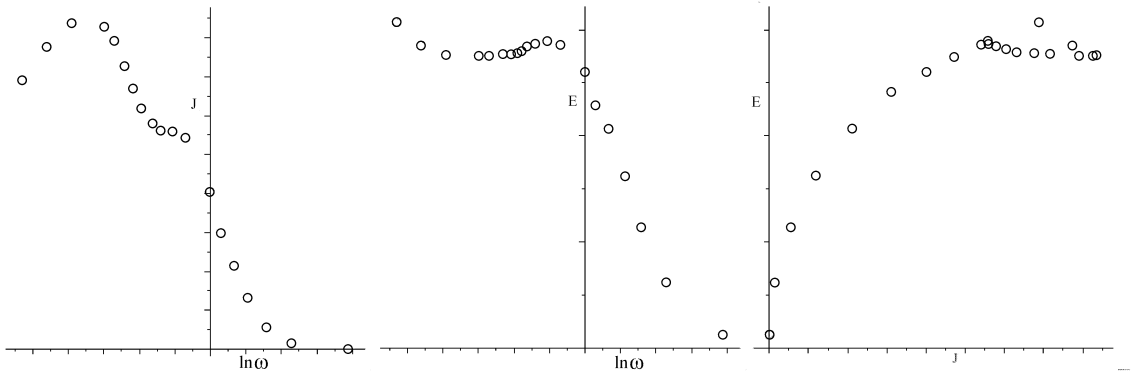}
 \end{center}
\caption{In order left to right for $p=3$ $b=10$: i) $J$ v. $\log \omega$; ii) $E$ v. $\log \omega$; iii) $E$ v. $J$. }
\end{figure}

As suggested by the boundary condition above where the angle depends on $\rho$ and $b$, one expects the $B$-field to introduce some asymmetry. It does this by tilting the string configuration. For constant $\omega$, the effect is captured in Fig. (\ref{P2b}). We note that in agreement with \cite{Arean:2005ar,Haque:2009hz}, the depth in the geometry probed by the string is roughly the same.

In plotting $E(\omega)$ versus $J(\omega)$, we observe a kink forming once one has left the $\omega >>1$ Regge regime Fig. (\ref{EJNC}). This is a natural expectation: as $J$ increases, the string endpoints separate, and we encounter non-commutative effects. 

\subsection{Non-extremal case}

\begin{figure}
\label{P2}
 \begin{center}
  \includegraphics[scale=.4]{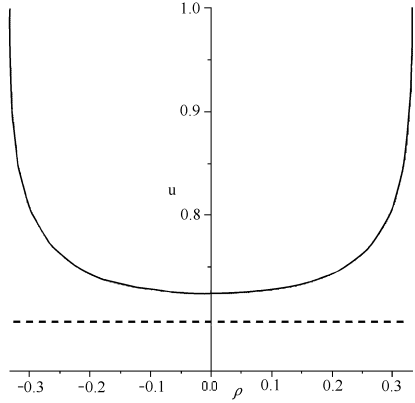}
 \end{center}
\caption{First node $n=0$ configuration for $p=2$ at $u_{T} = 0.7$.}
\end{figure}

For this case the action for this background (\ref{met2}) is 
\be
\label{act2}
S = -\frac{1}{2 \pi} \int d \tau d \sigma \left(\frac{u}{R}\right)^{\frac{7-p}{2}} \sqrt{(f-\omega^2 \rho^2)[ \rho'^2 + \left(\frac{R}{u}\right)^{7-p} \frac{u'^2}{f} ]},
\ee
while the energy and the angular momentum become:
\bea
E &=&  \frac{1}{2 \pi} \int d \sigma \left(\frac{u}{R}\right)^{\frac{7-p}{2}} f \sqrt{\frac{\rho'^2 + \left(\frac{R}{u}\right)^{7-p} \frac{u'^2}{f}}{f-\omega^2 \rho^2}}, \nn
J  &=&  \frac{\omega}{2 \pi} \int  d \sigma  \left(\frac{u}{R}\right)^{\frac{7-p}{2}} \rho^2 \sqrt{\frac{\rho'^2 + \left(\frac{R}{u}\right)^{7-p} \frac{u'^2}{f}}{f-\omega^2 \rho^2}}.
\eea

We choose the same boundary conditions as (\ref{bdy1}) with the strings ending orthogonal to the probe brane. For the non-extremal case, the presence of the horizon places a cut-off on the higher excited modes. Fig. (\ref{P2}) shows the only surviving configuration at $U_{T} = 0.7$ for $p=2$. The story is qualitatively the same in different dimensions. As the temperature is increased, the horizon advances and gradually suffocates the rotating strings. Any string that originally would have stretched into the space now occupied by the black hole will disappear leaving physically an open string split into two pieces. From the field theory point of view, this should correspond to string-splitting and the resulting melting of the positronium-like state. For a selection of works on this subject, see \cite{Mateos:2006nu, Peeters:2005fq}.  

\begin{figure}
\label{EJNE}
 \begin{center}
  \includegraphics[scale=.35]{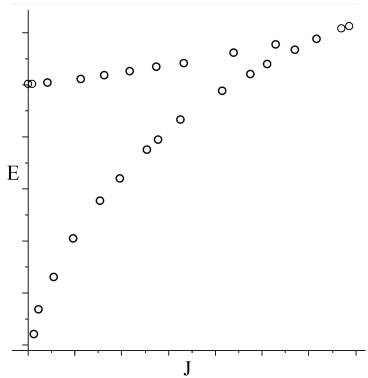}
 \end{center}
\caption{$E$ v. $J$ at high temperature ($u_{T} = 0.7$) for $p=3$.}
\end{figure}

For $E(\omega)$ versus $J(\omega)$, Fig. (\ref{EJNE}) we observe for high temperature that once beyond Regge limit of the graph, with decreasing $\omega$, the graph reaches a stage where $J$ does not increase with decreasing $\omega$ as was observed previously. Instead, with decreasing $\omega$, the graph, beyond this stage, changes direction drastically culminating in a second branch. This second branch is more energetic for a given value of $J$, so it is presumably unstable. It is natural to interpret the maximum value of $J$ as the point where the mesons dissociate and melt. For a Sakai-Sugimoto setup \cite{Sakai:2004cn}, a very similar maximum value of $J$ was noted in \cite{Peeters:2006iu}. 

\section*{{\large Acknowledgements}}
We are grateful to  Ki-Myeong Lee, Sun Min Park, Shahin Sheikh-Jabbari, Piljin Yi and Ignacio Zaballa for discussion.

\end{document}